\documentclass[preprint]{ptptex}

\usepackage{wrapft}

\usepackage{amsmath,amssymb,amscd,amsbsy,amsgen,amsopn,amstext,amsxtra}
\usepackage[mathscr]{eucal}

\newcommand{\iDsla}{iD\!\!\!\!/\,}
\newcommand{\itDsla}{i\tilde{D}\!\!\!\!/\,}

\title{%        %You can use \\ for explicit line-break
Atiyah-Singer Index Theorem in an $\boldsymbol{SO(3)}$ 
Yang-Mills-Higgs System 
and Derivation of a Charge Quantization Condition 
}
\author{%       %Use \scshape  for the family name
Shinichi \textsc{Deguchi} \footnote{E-mail:  deguchi@phys.cst.nihon-u.ac.jp}  
}

\inst{%         %Affiliation, neglected when [addenda] or [errata]
Institute of Quantum Science, College of Science and Technology, \\ 
Nihon University, Tokyo 101-8308, Japan
}

\recdate{%      %Editorial Office will fill in this.
May 23, 2007
}

\abst{%     
The Atiyah-Singer index theorem is generalized to a two-dimensional $SO(3)$  
Yang-Mills-Higgs (YMH) system.  
The generalized theorem is proven by using 
the heat kernel method and a nonlinear realization of $SU(2)$ gauge symmetry.  
This theorem is applied to the problem of deriving a charge quantization condition in the 
four-dimensional $SO(3)$ YMH system with non-Abelian monopoles. 
The resulting quantization condition, 
$eg=\frak{n}$~($\frak{n} \in \Bbb Z$), for an electric charge $e$ and a magnetic charge $g$   
is consistent with that found by Arafune, Freund and Goebel. 
It is shown that the integer $\frak{n}$ 
is half of the index of a Dirac operator.}

\begin{document}

\maketitle

\numberwithin{equation}{section}

%%%%%%%%%%%%%%%%%%%%%%%%%%%%%%%%%%%%%%%%%%%%
%SECTION 1

\section{Introduction}
The theory of magnetic monopoles has been studied   
by many people from various physical and mathematical points of view.\cite{Dir}\tocite{Shn} 
Although there is yet no experimental evidence of the existence of magnetic monopoles, 
it is believed that monopoles play crucial roles 
in long-standing problems in theoretical physics, 
such as the grand unification of forces and the confinement problem in quantum 
chromodynamics.

One of the most important consequences of the theory of monopoles is that electric charges  
are quantized in units that are inversely proportional to the magnetic charge of the monopole.  
Such an interesting role of monopoles was discovered by Dirac within the framework  
of quantum mechanics\cite{Dir}. In the natural units such that $\hbar=c=1$,  
the quantization condition shown by Dirac reads 
\begin{align}
eg=\frac{n}{2}\,, \quad n \in \Bbb Z \,, 
\label{1.1}
\end{align}
where $e$ and $g$ denote the electric and magnetic charges, respectively. 
Another charge quantization condition was discovered by Schwinger in his study of 
relativistic quantum electrodynamics with a magnetic charge\cite{Sch}. 
Schwinger's quantization condition reads 
\begin{align}
eg=n\,, \quad n \in \Bbb Z \,. 
\label{1.2}
\end{align}
The difference between these two conditions is essentially due to a difference in   
the string singularities of the gauge potentials adopted in the Dirac and Schwinger 
formalisms\cite{Fel}.

Recently, 
both the conditions (\ref{1.1}) and (\ref{1.2}) were derived   
in a unified manner by utilizing the Atiyah-Singer index theorem in two dimensions\cite{DK}. 
(For the Atiyah-Singer index theorem in any even number of dimensions, see, e.g.,  
Refs.~\citen{AS,ABP,NS,EGH,Gil,Nak,Ber,FS}.)  
The approach taken there is a second quantized approach in the sense that 
all the eigenfunctions of the Dirac operator are taken into account simultaneously.  
Unlike previous approaches, this approach requires neither classical notion 
for paths around a string singularity nor the concepts of patches and sections. 
To derive Eqs. (\ref{1.1}) and (\ref{1.2}), 
it is only necessary to solve a simple Dirac equation in two-dimensions and to formally 
count the number of zero-modes of the Dirac operator.  
Each of Eqs. (\ref{1.1}) and (\ref{1.2}) can be regarded as 
the necessary and sufficient condition that the Atiyah-Singer index theorem in two 
dimensions be valid for the $U(1)$ gauge theory with a monopole background.

The above-mentioned conditions are concerned with Abelian monopoles.  
In addition to Abelian monopoles, there also exist so-called non-Abelian monopoles. 
They are realized in some non-Abelian gauge theories 
as solitonic objects.\cite{tHo}\tocite{Shn} 
Non-Abelian monopoles have been studied since 't  Hooft and Polyakov 
independently discovered that a simultaneous system of field equations in 
the $SO(3)$ Yang-Mills-Higgs (YMH) theory admits a static solution representing  
monopoles with unit magnetic charge.\cite{tHo,Pol}   
Arafune, Freund and Goebel clarified the geometric origin of the conserved magnetic charge 
defined in the $SO(3)$ YMH theory,  and     
found the charge quantization condition valid in this theory,\cite{AFG} 
\begin{align}
eg=\frak{n}\,, \quad \frak{n} \in \Bbb Z \,. 
\label{1.3}
\end{align}
Here, $e$ is the Yang-Mills electric charge, and $g$ denotes the conserved magnetic charge.   
Arafune et al. demonstrated this condition by considering the homotopy class 
of a triplet of Higgs fields.  
Configurations of monopoles with non-minimal magnetic charge,  
$g=\frak{n}/e$ $(|\frak{n}|\geq 2)$,  were constructed 
by Bais using a method in which the Dirac monopole potential 
is embedded in the $SU(2)$ Lie algebra.\cite{Bai}

In this paper, we study an alternative approach to deriving Eq. (\ref{1.3}) 
using a generalization of the Atiyah-Singer index theorem in two dimensions. 
This approach is a non-Abelian analog of the approach taken 
in Ref.~\citen{DK} to derive Eqs. (\ref{1.1}) and (\ref{1.2}). 
In order to treat $SO(3)$ non-Abelian monopoles, we first consider a generalization of    
the Atiyah-Singer index theorem to a two-dimensional $SO(3)$ YMH system.  
Although the generalized theorem itself is valid  
for any two-dimensional manifold with spin structure,  
here we choose a sphere embedded in four-dimensional spacetime as the   
two-dimensional manifold. 
Then, taking the radius of the sphere to be infinitely large, 
we actually solve a massless Dirac equation on this sphere  
to find the zero-modes of the Dirac operator included in this equation.  
After examining the number of chirality zero-modes of the Dirac operator, 
it is shown that the generalized Atiyah-Singer index theorem leads to     
the charge quantization condition (\ref{1.3}).  
In this argument, the integer $\frak{n}$ is defined as half of the index 
of the Dirac operator.

This paper is organized as follows. 
Section 2 presents the Atiyah-Singer index theorem generalized to a two-dimensional $SO(3)$ 
YMH system. A simple proof of this theorem is also given there. 
In \S3, the generalized theorem is applied to deriving a charge quantization condition, 
which is shown to be identical to Eq. (\ref{1.3}). 
Section 4 is devoted to a summary and discussion. 
The appendix gives concrete forms of the $SO(3)$ monopole configurations   
and, with these configurations, illustrates 
the charge quantization condition derived in \S3.

%%%%%%%%%%%%%%%%%%%%%%%%%%%%%%%%%%%%%%%%%%%%
%SECTION 2

\section{Atiyah-Singer index theorem in an $SO(3)$ YMH system}

Let $\mathcal{M}$ be a compact, oriented, two-dimensional Riemannian manifold 
without boundary. 
Let $A$ be a hermitian Yang-Mills connection on $\mathcal{M}$ 
that takes values in the Lie algebra $\frak{su}(2)$ of the group $SU(2)$, 
and let $\varPhi$ be a hermitian scalar field on  
$\mathcal{M}$ that also takes values in $\frak{su}(2)$. 
Then $A$ and $\varPhi$ can be expanded as 
$A=A^{I}\tau_{I}\,$ $(I=1,2,3)$ 
and $\varPhi=\varPhi^{I}\tau_{I}$ in terms of the Pauli matrices $\tau_{I}$.  
The component fields $(A^{I}, \varPhi^{I})$ constitute an $SO(3)$ YMH system. 
Here, we impose the normalization condition 
${\rm tr}(\varPhi \varPhi)=2$, or equivalently 
$\sum_{I=1}^{3} (\varPhi^{I})^{2}=1$, 
without destroying the $SO(3)$ symmetry of the system;   
accordingly, $(\varPhi^{I})$ is treated as a normalized triplet of Higgs fields. 
Under this condition, it is possible to diagonalize $\varPhi$ in such a way that  
$v^{\dagger} \varPhi v=\tau_{3}$, with $v\in SU(2)$.  
Thus, $\varPhi$ can be represented as\footnote{The matrix $v=v(\varPhi)$  
is completely determined from $\varPhi$ up to a phase factor. 
Under the left action of $g \in SU(2)$, 
the matrix $v(\varPhi)$ transforms as  
$g v(\varPhi)=v(\varPhi^{\prime}) h$, with $h\in U(1)$. 
In this way, the $SU(2)$ gauge symmetry can be realized in a nonlinear manner 
with the aid of $v(\varPhi)$\cite{SS,Nie}.  
The matrices $\{ v(\varPhi)\}$ that correspond to the possible values of $\varPhi$
are sometimes referred to as {\em coset representatives}   
of the coset space $SU(2)/U(1)$.} 
\begin{align} 
\varPhi=v \tau_{3} v^{\dagger} . 
\label{1}
\end{align}
From $A$, $\varPhi$ and $\varPsi_{i}\equiv v \tau_{i} v^{\dagger}\,$ $(i=1,2)$,  
we define the {\em new} connection 
\begin{align}
A^{\perp} \equiv A- \frac{1}{2e} \epsilon_{ij3} {\rm tr}( \varPsi_i D \varPhi) \varPsi_j , 
\label{2}
\end{align}
where $D \varPhi \equiv d \varPhi -i(e/2)[A, \varPhi]$, and $e$ is an electric charge.  
Obviously, $A^{\perp}$ takes values in $\frak{su}(2)$.

Assuming that $\mathcal{M}$ possesses a spin structure, 
we consider a self-adjoint Dirac operator $\iDsla^{\perp}$ containing $A^{\perp}$  
instead of $A$.  
In terms of local coordinates 
$(q^\alpha)$ $(\alpha=1,2)$ on $\mathcal{M}$, the Dirac operator $\iDsla^{\perp}$ 
can be expressed as  
\begin{align} 
\iDsla^{\perp} \equiv i( \tau_0 \otimes \sigma_a) e_{a}{}^{\alpha} D^{\perp}_{\alpha} \,,
\label{3}
\end{align}
with 
\begin{align}
D^{\perp}_{\alpha} \equiv  
(\tau_0 \otimes \sigma_0) \partial_\alpha 
+ \frac{i}{2} \tau_0 \otimes (\omega_{\alpha} \sigma_{3}) 
-i\frac{e}{2} A^{\perp}_{\alpha} \otimes \sigma_0  \,.
\label{4}
\end{align}
Here $\partial_\alpha \equiv \partial/\partial q^\alpha$, 
$e_{a}{}^{\alpha}$~$(a=1,2)$ is an inverse zweibein on $\mathcal{M}$, and 
$\omega_{\alpha} $ is a spin connection in two dimensions. 
Both $\tau_0$ and $\sigma_0$ denote the $2\times 2$ unit matrices, while  
$\sigma_a$ and $\sigma_3$ denote the Pauli matrices which are understood as 
Dirac matrices in two dimensions. 
The symbol $\otimes$ stands for the tensor product of the $\tau$-matrices and the 
$\sigma$-matrices. It is obvious that 
the operator $\iDsla^{\perp}$ can be represented as a $4\times 4$ matrix.

Let $\varphi_{0, \nu_{t,s}}^{t,s}$ $(t,s=+,-\,;\,\nu_{t,s}=1,\ldots,\frak{n}_{t,s})$ 
be chirality zero-modes of $\iDsla^{\perp}$ characterized by 
\begin{align}
& \iDsla^{\perp} \varphi_{0, {\nu}_{t,s}}^{t,s}=0\,, 
\label{5} 
\\
& (\varPhi \otimes \sigma_0) \varphi_{0, {\nu}_{t,s}}^{t,s} 
=t  \varphi_{0, {\nu}_{t,s}}^{t,s} \,,  
\label{6} 
\\  
& (\tau_0 \otimes \sigma_3) \varphi_{0, {\nu}_{t,s}}^{t,s} 
=s  \varphi_{0, {\nu}_{t,s}}^{t,s} \,. 
\label{7}
\end{align}
Here, $\frak{n}_{t,s}$ denotes the number 
of chirality zero-modes specified by $(t,s)$. 
With Eqs. (\ref{1}) and (\ref{3}), we can verify that 
$[ \iDsla^{\perp}, \varPhi \otimes \sigma_0 ]
=(v \otimes \sigma_0) [ \itDsla^{\perp}, \tau_3 \otimes \sigma_0 ]
(v^{\dagger} \otimes \sigma_0)=0$, where  
$\itDsla^{\perp}$ is defined by replacing $A_{\alpha}^{\perp}$ 
contained in Eq. (\ref{3}) by 
$\Tilde{A}^{3}_{\alpha} \tau_3$, the third component of 
$\Tilde{A}_{\alpha}\equiv v^{\dagger} A_{\alpha} v +(2i/e) v^{\dagger} \partial_{\alpha} v
=\Tilde{A}_{\alpha}^{I} \tau_{I}$. 
Hence, Eqs. (\ref{5}) and (\ref{6}) can be satisfied simultaneously. 
Having defined the chirality zero-modes by Eqs. (\ref{5})--(\ref{7}), 
we can state the Atiyah-Singer index theorem generalized to a two-dimensional 
SO(3) YMH system:      
\begin{align}
\frak{n}_{++} -\frak{n}_{+-} -\frak{n}_{-+} +\frak{n}_{--} 
=\frac{e}{4\pi} \int_{\cal M} d^2 q  \, \varepsilon^{\alpha\beta} 
\mathcal{F}_{\alpha\beta} \,. 
\label{8}
\end{align}
Here, $\varepsilon^{\alpha\beta}$ $(\varepsilon^{12}=1)$ is the contravariant 
Levi-Civita tensor density in two dimensions, and 
$\mathcal{F}_{\alpha\beta}$ is the {\em 't Hooft tensor}\cite{tHo} in two dimensions,
\begin{align}
\mathcal{F}_{\alpha\beta}
&\equiv \frac{1}{2}\, {\rm tr} \bigg[\varPhi  
F_{\alpha\beta}+\frac{i}{2e} \varPhi
\big(D_{\alpha} \varPhi D_{\beta} \varPhi 
-D_{\beta} \varPhi D_{\alpha} \varPhi \big) \bigg] \,, 
\label{9}
\end{align}
with  
$F_{\alpha\beta} \equiv \partial_{\alpha}A_{\beta}-\partial_{\beta}A_{\alpha}
-i(e/2)[A_{\alpha}, A_{\beta}]$.\footnote{
In component form,  $F_{\alpha\beta}$ and $D_{\alpha}\varPhi$ 
can be expressed as 
$F_{\alpha\beta}^{I}=\partial_{\alpha}A_{\beta}^{I}-\partial_{\beta}A_{\alpha}^{I}
+e \epsilon^{IJK} A_{\alpha}^{J} A_{\beta}^{K}$ and 
$D_{\alpha}\varPhi^{I}=\partial_{\alpha}\varPhi^{I} 
+e \epsilon^{IJK} A_{\alpha}^{J} \varPhi^{K}$. 
From these, we see that the fundamental electric charge, 
or the gauge coupling constant, in the present $SO(3)$ YMH system is $e$, not $e/2$. 
In terms of the component fields, the 't Hooft tensor (\ref{9}) is written   
$\mathcal{F}_{\alpha\beta}=\varPhi^{I}F_{\alpha\beta}^{I}
-(1/e)\epsilon_{IJK}\varPhi^{I} D_{\alpha}\varPhi^{J} D_{\beta}\varPhi^{K}$. 
This often appears in the literature on non-Abelian monopoles.
\cite{tHo,AFG}\tocite{Shn}}
The left-hand side of Eq. (\ref{8}) is referred to as the Atiyah-Singer index of the Dirac operator 
$\iDsla^{\perp}$. In the remaining part of this section,  
we prove the Atiyah-Singer index theorem (\ref{8}).

To this end, we first consider the eigenvalue equation 
\begin{align}
\iDsla^{\perp} \varphi_N (q)=\lambda_n  \varphi_N (q) \,, 
\label{10}
\end{align}
with an eigenvalue $\lambda_{n}$ and an eigenfunction $\varphi_{N}$. 
Here, $N$ is a collective index, $N=(n, \nu)$,   
where $\nu$ is a label that distinguishes between the degenerate eigenfunctions 
corresponding to $\lambda_n$.  
The eigenfunction $\varphi_{N}$ is assumed to be sufficiently smooth that  
$[ \partial_\alpha, \partial_\beta ]\varphi_{N}=0$ holds. 
Because $\iDsla^{\perp}$ is self-adjoint, the eigenvalue $\lambda_{n}$ is purely real,  
and the eigenfunctions $\{ \varphi_N \}$ can be assumed to form a complete 
orthonormal set.  Now we evaluate the function 
\begin{align} 
\mathcal{A}_{\rm reg} (q)  
& \equiv 
\lim_{\varsigma \searrow 0} 
\sum_{N} \varphi^{\dagger}_N (q)
 (\varPhi \otimes \sigma_3) e^{-\varsigma \lambda_n^2} \varphi_N (q)  
\nonumber \\ 
& =\lim_{\varsigma\searrow 0} 
\sum_{N} \varphi^{\dagger}_N (q) 
(\varPhi \otimes \sigma_3)  
\exp\!\big[-\varsigma (\iDsla^{\perp})^2 \big]  \varphi_N (q) 
\nonumber \\ 
& =\lim_{\varsigma \searrow 0} \lim_{q'\rightarrow q}
 {\rm Tr}\big( (\varPhi \otimes \sigma_3) G^{\perp} (q,q', \varsigma) \big) \,, 
\label{11} 
\end{align}
with 
\begin{align}
G^{\perp} (q,q', \varsigma)\equiv \sum_{N} 
\big\{ \exp \big[-\varsigma (\iDsla^{\perp})^2 \big] 
\varphi_N (q) \big\} \varphi^{\dagger}_N (q') \,.
\label{12}
\end{align}
Here, ^^ ^^ Tr" represents the trace taken over both the 
$\tau$- and $\sigma$-matrices.  Following the procedure used in Ref.~\citen{DK},  
which is based on the heat kernel method\cite{Ber, Par, BB, Avr, Vas}, 
we can calculate the two-point function $G^{\perp}(q,q',\varsigma)$  
in the limit $q'\rightarrow q$. 
From this calculation, we obtain, for $0< \varsigma \ll 1$, 
\begin{align}
& \lim_{q' \rightarrow q} G^{\perp} (q,q',\varsigma) 
\nonumber \\
&\, = \frac{1}{4\pi \varsigma} \tau_0 \otimes \sigma_0
-{1\over48\pi} \tau_0 \otimes (R\sigma_0)
+\frac{e}{16\pi}  \epsilon^{\alpha\beta} F_{\alpha\beta}^{\perp} \otimes \sigma_3  
+O(\varsigma)  \,,
\label{13}
\end{align}
where 
$\epsilon^{\alpha\beta} \equiv |\det(e_{a}{}^{\alpha})| \varepsilon^{\alpha\beta}$, 
$F^{\perp}_{\alpha\beta} \equiv 
\partial_{\alpha}A^{\perp}_{\beta}-\partial_{\beta}A^{\perp}_{\alpha}
-i(e/2)[A^{\perp}_{\alpha}, A^{\perp}_{\beta}]$, and $R$ is the scalar curvature of  
$\mathcal{M}$. 
Inserting Eq. (\ref{13}) into Eq. (\ref{11}) and evaluating the trace over the  
$\sigma$-matrices lead to  
\begin{align} 
\mathcal{A}_{\rm reg} (q) =\frac{e}{8\pi}  {\rm tr} 
\big(\varPhi  \epsilon^{\alpha\beta} F^{\perp}_{\alpha\beta} \big) \,, 
\label{14}
\end{align}
where the trace over the $\tau$-matrices remains.

Next, we consider a (generalized) chiral decomposition of $\varphi_N$, 
\begin{align}
\varphi^{t,s}_{N} \equiv \frac{1}{4}
\{ (\tau_0 +t\varPhi) \otimes (\sigma_0 +s\sigma_3) \} \varphi_{N} , 
\label{15}
\end{align}
where $t, s=+, -$. 
Because $\varPhi^2=(v\tau_3 v^{\dagger})^2 =\tau_0$ 
and $\sigma_3^2=\sigma_0$,  it is easy to see 
that the components $\varphi^{t,s}_{N}$ satisfy the eigenvalue equations 
\begin{align}
& (\varPhi \otimes \sigma_0) \varphi^{t,s}_{N}  =t  \varphi^{t,s}_{N} \,, 
\label{16}
\\ 
&(\tau_0  \otimes \sigma_3) \varphi^{t,s}_{N} =s  \varphi^{t,s}_{N} \,. 
\label{17}
\end{align}
Furthermore, $\varphi^{t,s}_{N}$ satisfy the orthogonality relations  
$\varphi_{N}^{t,s\,\dagger} \varphi_{N'}^{-t,s'} 
=\varphi_{N}^{t,s\,\dagger}  \varphi_{N'}^{t',-s}=0$. 
In terms of  $\varphi^{t,s}_{N}$, Eq. (\ref{10}) can be written 
\begin{align}
\iDsla^{\perp} \varphi^{t,s}_{N} (q)=\lambda_n  \varphi^{t,-s}_{N} (q) \,. 
\label{18}
\end{align}
Here we assume that $\lambda_0 =0$. Thereby, 
the corresponding eigenfunctions $\varphi_{0, {\nu}_{t,s}}^{t,s}$ are treated as 
the chirality zero-modes of $\iDsla^{\perp}$, and 
Eqs. (\ref{5}), (\ref{6}) and (\ref{7}) are understood as 
Eqs. (\ref{18}), (\ref{16}) and (\ref{17}) in the case $n=0$, respectively.  
Equation (\ref{18}) shows that  
when $n\neq0$, there is a one-to-one correspondence between 
$\varphi^{t,+}_{N}$ and $\varphi^{t,-}_{N}$. Consequently, it follows that 
the number of elements of $\{\varphi_{N}^{t,+}\}_{n\neq0}$ is equal to 
the number of elements of $\{\varphi_{N}^{t,-}\}_{n\neq0}$. 
Also, when $n\neq 0$, it can be proved using Eq. (\ref{18}) that  
\begin{align}
\int_\mathcal{M} d^2 q \sqrt{\frak{g}(q)}\,  
\varphi^{t,+\,\dagger}_{N} (q) \varphi^{t,+}_{N} (q) 
=\int_\mathcal{M} d^2 q \sqrt{\frak{g}(q)}\,  
\varphi^{t,-\,\dagger}_{N} (q) \varphi^{t,-}_{N} (q) \, , 
\label{19}
\end{align}
with $\sqrt{\frak{g}} \equiv |\det(e_{a}{}^{\alpha})|^{-1}$.  
Since the zero-modes $\varphi_{0, {\nu}_{t,s}}^{t,s}$ are eigenfunctions of 
$\iDsla^{\perp}$, the set $\big\{ \varphi_{0, {\nu}_{t,s}}^{t,s} \big\}$ 
satisfying the orthonormality condition 
\begin{align}
\int_\mathcal{M} d^2 q \sqrt{\frak{g}(q)}\, 
\varphi_{0, {\nu}_{t,s}}^{t,s \,\dagger} (q)  
\varphi_{0, {\nu}_{t',s'}}^{t',s'} (q) 
=\delta_{tt'} \delta_{ss'} \delta_{{\nu}_{t,s}, {\nu}_{t',s'}}  
\label{20}
\end{align} 
can be taken as a subset of the orthonormal set $\{\varphi_N \}$. 
From Eq. (\ref{15}), we see that $\varphi_{N}=\sum_{t,s} \varphi_{N}^{t,s}$. 
Substituting this into the first line of Eq. (\ref{11}), and using 
$ (\varPhi \otimes \sigma_3) \varphi_{N}^{t,s} 
=(\varPhi \otimes \sigma_0) (\tau_{0} \otimes \sigma_3) \varphi_{N}^{t,s} 
=ts  \varphi_{N}^{t,s}$ and 
Eqs. (\ref{19}) and (\ref{20}),  
we have  
\begin{align}
& \int_\mathcal{M} d^2 q \sqrt{\frak{g}(q)}\, \mathcal{A}_{\rm reg} (q) 
\nonumber 
\\
& = \lim_{\varsigma \searrow 0}
\sum_{N} e^{-\varsigma \lambda_{n}^2} \sum_{t,s} t s 
\int_\mathcal{M} d^2 q \sqrt{\frak{g}(q)}\,  
\varphi_{N}^{t,s\,\dagger} (q) \varphi_{N}^{t,s} (q) 
\nonumber 
\\ 
& =
\sum_{t,s} t s \sum_{ {\nu}_{t,s} } \int_\mathcal{M} d^2 q \sqrt{\frak{g}(q)}\,  
\varphi_{0, {\nu}_{t,s}}^{t,s\,\dagger} (q)  \varphi_{0, {\nu}_{t,s}}^{t,s} (q) 
=\sum_{t,s} t s  \frak{n}_{t,s} \,.  
\label{21}
\end{align} 
Combining Eqs. (\ref{14}) and (\ref{21}) yields  
\begin{align}
\frak{n}_{++} -\frak{n}_{+-} -\frak{n}_{-+} +\frak{n}_{--} 
=\frac{e}{8\pi} \int_{\cal M} d^2 q  \,  
{\rm tr} \big(\varPhi \varepsilon^{\alpha\beta} F^{\perp}_{\alpha\beta} \big)  \,. 
\label{22}
\end{align}

Now, with Eq. (\ref{1}), it is easy to show 
$ v^{\dagger} (D \varPhi) v=e \epsilon_{3ij} \tau_{i} \tilde{A}^{j}$, or 
\begin{align}
D \varPhi =e \epsilon_{3ij} \varPsi_{i} \tilde{A}^{j} , 
\label{23}
\end{align}
where  
\begin{align}
\tilde{A}\equiv v^{\dagger} Av+\frac{2i}{e} v^{\dagger} dv =\tilde{A}^{I} \tau_{I} \,. 
\label{23.5}
\end{align}
Substituting Eq. (\ref{23}) into Eq. (\ref{2}), we can express $A^{\perp}$ as 
$A^{\perp}= v \tilde{A}^{3} \tau_3 v^{\dagger} +(2i/e) v dv^{\dagger}$. 
Since this is just a gauge transformation of $\tilde{A}^{3} \tau_3$, 
the field strength $F_{\alpha\beta}^{\perp}$ can be written as  
$F_{\alpha\beta}^{\perp} =v\tilde{F}_{\alpha\beta}^{\perp 3} \tau_3 v^{\dagger}$, 
with $\tilde{F}_{\alpha\beta}^{\perp 3} 
\equiv \partial_{\alpha} \tilde{A}_{\beta}^{3} -\partial_{\beta} \tilde{A}_{\alpha}^{3}$. 
Then, it follows that 
\begin{align}
{\rm tr} \big( \varPhi F_{\alpha\beta}^{\perp} \big)
=2\tilde{F}_{\alpha\beta}^{\perp 3} \, . 
\label{24}
\end{align}
Using Eq. (\ref{23}),  it is readily shown that 
${\rm tr}(\varPhi D\varPhi D\varPhi)=e^2 {\rm tr} (\tau_3 \tilde{A}^{2} )$.  
Also, expressing the field strength $F_{\alpha\beta}$ as 
$F_{\alpha\beta}=v \tilde{F}_{\alpha\beta} v^{\dagger}$, 
with $\tilde{F}_{\alpha\beta} \equiv 
\partial_{\alpha} \tilde{A}_{\beta}-\partial_{\beta} \tilde{A}_{\alpha}
-i(e/2)[\Tilde{A}_{\alpha}, \tilde{A}_{\beta}]$, we see that 
${\rm tr}(\varPhi F_{\alpha\beta})={\rm tr} (\tau_3 \tilde{F}_{\alpha\beta} )$. 
Then, we can write the 't Hooft tensor (\ref{9}) as   
\begin{align}
\mathcal{F}_{\alpha\beta}
&=\frac{1}{2} \,{\rm tr} \bigg[ \tau_3 \tilde{F}_{\alpha\beta}
+\frac{i}{2e} e^2 \tau_{3} 
\big (\tilde{A}_{\alpha} \tilde{A}_{\beta} -\tilde{A}_{\beta} \tilde{A}_{\alpha} \big) \bigg]
\nonumber 
\\ 
&=\frac{1}{2}\, {\rm tr} \big[ \tau_3 
\big( \partial_{\alpha} \Tilde{A}_{\beta}-\partial_{\beta} \Tilde{A}_{\alpha} \big) \big] 
=\Tilde{F}_{\alpha\beta}^{\perp 3} \,, 
\label{25}
\end{align}
which makes it clear that the 't Hooft tensor is indeed an Abelian field strength. 
Combining Eqs. (\ref{24}) and (\ref{25}) yields         
${\rm tr}\big( \varPhi F_{\alpha\beta}^{\perp} \big) =2\mathcal{F}_{\alpha\beta}$.  
Inserting this into Eq. (\ref{22}) leads to Eq. (\ref{8}). 
Thus, the Atiyah-Singer index theorem in an $SO(3)$ YMH system,  
Eq. (\ref{8}), is proved. 

%%%%%%%%%%%%%%%%%%%%%%%%%%%%%%

\section{Derivation of a charge quantization condition} 

In this section, we derive a charge quantization condition in  
the {\em static} $SO(3)$ YMH system in four-dimensional spacetime,  
$M^4$, by utilizing the Atiyah-Singer index theorem (\ref{8}).  
For this purpose,  we choose a sphere $S^2_R$ of radius $R$ 
embedded in $M^4$ as the two-dimensional manifold $\mathcal{M}$. 
To derive the correct charge quantization condition using Eq. (\ref{8}), 
we need to examine its left-hand side in detail,  
showing relations valid among the numbers 
$\frak{n}_{t,s}$ ($t,s=+,-$). 
These relations are beyond the Atiyah-Singer index theorem and 
can be found only by solving the Dirac equation (\ref{5})  
in the case $\mathcal{M}=S^2_R$ with $R\rightarrow \infty$. 
For this reason, 
we actually solve it in this section by carrying out appropriate 
gauge transformations so that Eq. (\ref{5}) takes a simple form.   
Then we show relations valid among $\frak{n}_{t,s}$ 
and derive a charge quantization condition using these relations and 
the Atiyah-Singer index theorem.

Having chosen the sphere $S^2_R$ as $\mathcal{M}$, it is natural for us to proceed  
with the study using spherical coordinates. 
In terms of spherical coordinates, $(q^1, q^2)=(\theta, \phi)\,$  
$(0\leq \theta \leq \pi,\, 0\leq \phi < 2\pi)$, on $S^{2}_R$,  
the diagonalized inverse zweibein $e_{a}{}^{\alpha}$ takes the form 
$(e_{a}{}^{\alpha})=\mathrm{diag}(R^{-1},\,R^{-1} \sin^{-1}\theta)$.  
The associated spin connection $\omega_{\alpha} $ is found to be  
$\omega_{\alpha}= -\delta_{\alpha 2} \cos\theta$\cite{DK}. 
We can regard the 't Hooft tensor $\mathcal{F}_{\alpha\beta}$ as the radial component 
of the magnetic field at $S^2_R$. 
In accordance with the theory of non-Abelian monopoles,  
the conserved magnetic charge in the $SO(3)$ YMH system is given by\cite{AFG,Shn}
\begin{align}
g=\frac{1}{4\pi} 
\int_{S^2_R} d^2 q  \,\frac{1}{2} \varepsilon^{\alpha\beta}  
\mathcal{F}_{\alpha\beta}
= \frac{1}{4\pi} 
\int_{S^2_R} \mathcal{F} \,, 
\label{3.1}
\end{align}
where the integral is evaluated in the limit $R\rightarrow \infty$. 
In this limit, 
the right-hand side of Eq. (\ref{8}) with $\mathcal{M}=S^2_R$ becomes $2eg$.

In order to make the left-hand side of Eq. (\ref{8})  
clearer in the case $\mathcal{M}=S^2_R$ with $R\rightarrow \infty$,   
we first rewrite Eq. (\ref{5}) as 
\begin{align}
\itDsla^{\perp} \tilde{\varphi}_{0}^{t,s}=0 \,, 
\label{3.2}
\end{align}
where 
\begin{align}
\tilde{\varphi}_{0}^{t,s} \equiv (v^{\dagger} \otimes \sigma_0) \varphi_{0}^{t,s} , 
\label{3.3}
\end{align}
and $\itDsla^{\perp}$ is defined by replacing $A_{\alpha}^{\perp}$ 
contained in Eq. (\ref{3}) by $\Tilde{A}^{3}_{\alpha} \tau_3$.  
(Here, the label $\nu_{t,s}$ is omitted for conciseness.)  
The transformation 
$(\varphi_{0}^{t,s}, A_{\alpha}^{\perp}) \rightarrow 
(\tilde{\varphi}_{0}^{t,s}, \Tilde{A}^{3}_{\alpha} \tau_3)$ 
is simply a gauge transformation. 
The zero-mode  $\tilde{\varphi}_{0}^{t,s}$ satisfies Eq. (\ref{7}) 
and $(\tau_3 \otimes \sigma_0) \tilde{\varphi}_{0}^{t,s} 
=t  \tilde{\varphi}_{0}^{t,s}$ instead of Eq. (\ref{6}). 
This implies that only the $(t,s)$-component of 
the four-column vector  
$\tilde{\varphi}_{0}^{t,s}$ remains non-vanishing, and thus   
$\tilde{\varphi}_{0}^{t,s}$ is expressed in component form as 
\begin{align}
{\tilde{\varphi}_{0}^{t,s}}{}_{t',s'}
=\delta^{t}{}_{t'} \delta^{s}{}_{s'} \tilde{u}^{t,s} . 
\label{3.3.1}
\end{align}
Here, $\tilde{u}^{t,s}$ is a function of $(\theta, \phi)$.  
In terms of $\tilde{u}^{t,s}$, the Dirac equation (\ref{3.2}) is written 
\begin{align}
\bigg[\, \frac{\partial}{\partial\theta} -it \frac{e}{2} \tilde{A}_{1}^{3}  +\frac{1}{2} \cot\theta 
+\frac{is}{\sin \theta} \bigg( \frac{\partial}{\partial \phi} -it \frac{e}{2} \tilde{A}_{2}^{3} \bigg) 
\bigg] \tilde{u}^{t,s} =0 \,. 
\label{3.4}
\end{align}
Now, consider the gauge transformation 
\begin{subequations}
\label{3.5}
\begin{align}
\tilde{u}^{t,s} &\longrightarrow \hat{u}^{t,s} \equiv \exp \!\bigg[ -it \frac{e}{2} \int^{\theta}_{0} 
\tilde{A}_{1}^{3}(\theta^{\prime},\phi) d\theta^{\prime} \bigg] \tilde{u}^{t,s} , 
\label{3.5a}
\\
\tilde{A}_{1}^{3} &\longrightarrow \hat{A}_{1}^{3} \equiv  0 \,,
\label{3.5b}
\\
\tilde{A}_{2}^{3} &\longrightarrow 
\hat{A}_{2}^{3} \equiv \tilde{A}_{2}^{3} 
-\frac{\partial}{\partial \phi} \int^{\theta}_{0} \tilde{A}_{1}^{3}(\theta^{\prime},\phi) 
d\theta^{\prime} \,, 
\label{3.5c}
\end{align}
\end{subequations}
which, of course, leaves $\tilde{F}_{12}^{\perp 3} 
= \partial_{1} \tilde{A}_{2}^{3} -\partial_{2} \tilde{A}_{1}^{3}$ invariant. 
Applying the gauge transformation (\ref{3.5}) to Eq. (\ref{3.4}),  
we can simplify it to     
\begin{align}
\bigg[\, \frac{\partial}{\partial\theta} +\frac{1}{2} \cot\theta
+\frac{is}{\sin \theta} \bigg( \frac{\partial}{\partial \phi} -it \frac{e}{2} \hat{A}_{2}^{3} \bigg) 
\bigg] \hat{u}^{t,s} =0 \,.
\label{3.6}
\end{align}
It should be noted that Eq. (\ref{3.5a}) 
is merely a regular phase transformation, because, unlike the azimuthal angle 
$\phi$, the polar angle $\theta$ is unrelated to winding of a closed path around an axis. 
For this reason, there is no essential difference between Eqs. (\ref{3.4}) and  
(\ref{3.6}), and the number of regular solutions of Eq. (\ref{3.4}) is equal to  
that of Eq. (\ref{3.6}). In the following, we treat Eq. (\ref{3.6}) to examine the 
number of chirality zero-modes of $\itDsla^{\perp}$. 
Because $\tilde{\varphi}_{0}^{t,s}$ is a spinor field, it has to change sign under a $2\pi$ 
rotation in $\phi$.  This condition and the single-valuedness of $\tilde{A}_{1}^{3}$ 
under a $2\pi$ rotation in $\phi$ lead to the anti-periodicity condition 
$\hat{u}^{t,s} (\theta, \phi+2\pi)=-\hat{u}^{t,s} (\theta, \phi)$ via 
Eqs. (\ref{3.3.1}) and (\ref{3.5a}). 
Accordingly, $\hat{u}^{t,s} (\theta, \phi)$ can be expressed as the Fourier series 
\begin{align}
\hat{u}^{t,s} (\theta, \phi) =\frac{1}{\sqrt{2\pi}}  
\sum_{m \in {\Bbb Z}+{\frac{1}{2}}} \hat{v}^{t,s}_{m}(\theta) e^{i m \phi} \,. 
\label{3.7}
\end{align} 
Substituting Eq. (\ref{3.7}) into Eq. (\ref{3.6}) and using the orthonormality relation 
\begin{align}
\int_{0}^{2\pi} \frac{d\phi}{2\pi}\,  e^{i (m-m^{\prime}) \phi} 
=\delta_{m, m^{\prime}} \,, \quad \; 
m, m^{\prime} \in {\Bbb Z}+{\frac{1}{2}} \,,
\label{3.8}
\end{align}
we obtain 
\begin{align}
\bigg( \frac{d}{d\theta} +\frac{1}{2} \cot\theta -\frac{s m}{\sin \theta} \bigg) 
\hat{v}^{t,s}_{m} 
+\frac{ste}{2\sin\theta} 
\sum_{m^{\prime}\in {\Bbb Z}+{\frac{1}{2}}} 
\int^{2\pi}_{0} \frac{d\phi}{2\pi}\, \hat{A}_{2}^{3} \, 
e^{i (m^{\prime}-m) \phi} \,\hat{v}^{t,s}_{m^{\prime}} =0 \,. 
\label{3.9}
\end{align}

Assume here that $\hat{A}^3_2$ is independent of $\phi$. Then, 
noting that $\tilde{F}_{12}^{\perp 3}$ can be written as 
$\tilde{F}_{12}^{\perp 3}=\partial_{1} \hat{A}_{2}^{3}$ by using Eq. (\ref{3.5c}),  
we see that $\tilde{F}_{12}^{\perp 3}$ is also independent of $\phi$ and 
depends only on $\theta$.  This condition is actually realized on the sphere 
$S_{\rm R}^{2}$ in the limit $R\rightarrow\infty$, 
because the magnetic field at $S_{R}^{2}$ becomes spherically symmetric  
as $R$ increases to infinity. 
(The condition that $\tilde{F}_{12}^{\perp 3}$ depends only on $\theta$ holds 
in more general situations in which the magnetic field at $S_{R}^{2}$ is axially symmetric.)  
Equation (\ref{3.9}) now reads  
\begin{align}
\bigg( \frac{d}{d\theta} +\frac{1}{2} \cot\theta -\frac{s m}{\sin \theta} 
+\frac{ste}{2\sin \theta} \hat{A}^3_2 (\theta) \bigg) 
\hat{v}^{t,s}_{m} =0 \,, 
\label{3.10}
\end{align}
whose solution is readily found to be 
\begin{align}
\hat{v}^{t,s}_{m}(\theta)=
{c}_{m}^{t,s}
\biggl(\sin\frac{\theta}{2}\biggr)^{\!sm-\frac{1}{2}} 
\biggl(\cos\frac{\theta}{2}\biggr)^{\!-sm-\frac{1}{2}}  
\exp\! \bigg[ -ste \int^{\theta}_{\theta_{0}} d\theta^{\prime}\, 
\frac{\hat{A}^3_2(\theta^{\prime})}{2\sin \theta^{\prime}} \bigg]  \,,
\label{3.11}
\end{align}
with ${c}_{m}^{t,s}$ being an appropriate constant. 
Here, we choose $c_{m}^{t,s}$ to be a normalization constant, 
if $\hat{v}^{t,s}_{m}(\theta)$ is regular, and hence normalizable,  
on the interval $0\leq \theta \leq \pi$.

Suppose that $st$ in Eq. (\ref{3.11}) is fixed, for instance, as $st=+$.  
Then, $\hat{v}^{++}_{m}=\hat{v}^{--}_{-m}$ is valid   
with the choice ${c}^{++}_{m}={c}^{--}_{-m}$, and hence 
the number of the regular solutions $\{ \hat{v}^{++}_{m} \}_{m\in  {\Bbb Z}^{\prime}+1/2}$
is equal to that of the regular solutions 
$\{ \hat{v}^{--}_{-m} \}_{m\in  {\Bbb Z}^{\prime}+1/2}$.  We simply express this fact as  
${\sharp} \{ \hat{v}^{++}_{m} \}_{m\in  {\Bbb Z}^{\prime}+1/2}
={\sharp} \{ \hat{v}^{--}_{-m} \}_{m\in  {\Bbb Z}^{\prime}+1/2}$. 
Here, ${\Bbb Z}^{\prime}$ denotes an appropriate subset of the set of integers ${\Bbb Z}$,  
which is found by examining the regularity of the solution (\ref{3.11}). 
(The symbol ${\sharp}\{\ast\}$ stands for the number of elements contained in $\{\ast\}$.)    
The fundamental set of solutions for Eq. (\ref{3.6}) is given by  
$\{ \hat{u}^{t,s}_{m} \}_{m\in  {\Bbb Z}+1/2}$, with 
$\hat{u}^{t,s}_{m} (\theta, \phi) \equiv (2\pi)^{-1/2} \hat{v}^{t,s}_{m}(\theta) e^{im \phi}$,  
and it is obvious that ${\sharp} \{ \hat{u}^{++}_{m} \}_{m\in  {\Bbb Z}^{\prime}+1/2}
={\sharp} \{ \hat{u}^{--}_{-m} \}_{m\in  {\Bbb Z}^{\prime}+1/2}$.  
The set $\{ \hat{u}^{t,s}_{m} \}_{m\in  {\Bbb Z}+1/2}$ yields the fundamental set of solutions 
for Eq. (\ref{3.4}), i.e. $\{ \tilde{u}^{t,s}_{m}\}_{m\in  {\Bbb Z}+1/2}$ 
with $\tilde{u}^{t,s}_{m} \equiv \exp \!\big[\,it(e/2) \int^{\theta}_{0} 
\tilde{A}_{1}^{3} d\theta^{\prime} \big] \hat{u}^{t,s}_{m}$, via Eq. (\ref{3.5a}).  
Because Eq. (\ref{3.5a}) is a regular phase transformation, as mentioned above,  
it is guaranteed that ${\sharp} \{ \tilde{u}^{++}_{m} \}_{m\in  {\Bbb Z}^{\prime}+1/2}
={\sharp} \{ \tilde{u}^{--}_{-m} \}_{m\in  {\Bbb Z}^{\prime}+1/2}$. 
Recalling that the regular solutions of Eq. (\ref{3.4}) lead to the zero-modes of 
$\itDsla^{\perp}$ in such a way that 
${\tilde{\varphi}_{0,m}^{t,s}}{}_{\,t',s'}=\delta^{t}{}_{t'} \delta^{s}{}_{s'} \tilde{u}^{t,s}_{m}$, 
we see that ${\sharp} \{ \tilde{\varphi}^{++}_{0,m} \}_{m\in  {\Bbb Z}^{\prime}+1/2}
={\sharp} \{ \tilde{\varphi}^{--}_{0,-m} \}_{m\in  {\Bbb Z}^{\prime}+1/2}$. 
The zero-modes of $\iDsla^{\perp}$ are connected with 
those of $\itDsla^{\perp}$ by the unitary transformation (\ref{3.3}): 
$\varphi_{0,m}^{t,s} = (v \otimes \sigma_0) \tilde{\varphi}_{0,m}^{t,s}$. 
Hence, it follows that 
${\sharp} \{ {\varphi}^{++}_{0,m} \}_{m\in  {\Bbb Z}^{\prime}+1/2}
={\sharp} \{ {\varphi}^{--}_{0,-m} \}_{m\in  {\Bbb Z}^{\prime}+1/2}$, or simply 
\begin{align}
\frak{n}_{++}=\frak{n}_{--} \,, 
\label{3.12}
\end{align}
where $\frak{n}_{t,s}$ for $ts=+$ is given by  
$\frak{n}_{t,s}={\sharp} \{ {\varphi}^{t,s}_{0,sm} \}_{m\in  {\Bbb Z}^{\prime}+1/2}$. 
If $st$ in Eq. (\ref{3.11}) is fixed as $st=-$,  then $\hat{v}^{+-}_{m}=\hat{v}^{-+}_{-m}$ 
is valid with the choice ${c}^{+-}_{m}={c}^{-+}_{-m}$.   
Following the same procedure as in the case $st=+$, 
we can show that ${\sharp} \{ {\varphi}^{+-}_{0,m} \}_{m\in  {\Bbb Z}^{\prime\prime}+1/2}
={\sharp} \{ {\varphi}^{-+}_{0,-m} \}_{m\in  {\Bbb Z}^{\prime\prime}+1/2}$.  
Here, ${\Bbb Z}^{\prime\prime}$ denotes an appropriate subset of ${\Bbb Z}$, 
which is different from ${\Bbb Z}^{\prime}$ in general. 
This relation for the numbers of chirality zero-modes is simply written  
\begin{align}
\frak{n}_{+-}=\frak{n}_{-+} \,, 
\label{3.13}
\end{align}
where $\frak{n}_{t,s}$ for $ts=-$ is given by  
$\frak{n}_{t,s}={\sharp} \{ {\varphi}^{t,s}_{0,-sm} \}_{m\in  {\Bbb Z}^{\prime\prime}+1/2}$.  
Using Eqs. (\ref{3.12}) and (\ref{3.13}), the left-hand side of  Eq. (\ref{8}) with  
$\mathcal{M}=S^2_R$ can be written as $2(\frak{n}_{++}-\frak{n}_{+-})$,  
at least in the limit $R\rightarrow \infty$.

As mentioned under Eq. (\ref{3.1}), the right-hand side of  Eq. (\ref{8}) with 
$\mathcal{M}=S^2_R$ becomes $2eg$ in the limit $R\rightarrow \infty$. 
Thus, in the present case,  Eq. (\ref{8}) reduces to
\begin{align}
\frak{n}_{++}-\frak{n}_{+-}=eg \,. 
\label{3.14}
\end{align}
The left-hand side of Eq. (\ref{3.14}) is just the difference between the numbers of  
positive ($s=+$) and negative ($s=-$) chirality zero-modes having the common signature $t=+$.   
The difference $\frak{n}_{++}-\frak{n}_{+-}$  is, of course, an integer, 
and by setting $\frak{n}=\frak{n}_{++}-\frak{n}_{+-}$, Eq. (\ref{3.14}) can be expressed as 
\begin{align}
eg=\frak{n} \,,  \quad \frak{n} \in \Bbb Z \,.
\label{3.15}
\end{align}
This is precisely the charge quantization condition (\ref{1.3}), 
although the integer $\frak{n}$ here 
has its own meaning. 
Thus, we have derived a correct charge quantization condition by utilizing  
the Atiyah-Singer index theorem (\ref{8}).

\section{Summary and discussion} 
In this paper, we first considered a generalization of the Atiyah-Singer index theorem 
to a two-dimensional  $SO(3)$ YMH system.  
The generalized theorem (\ref{8}) was proven by using 
the heat kernel method and a nonlinear realization of the $SU(2)$ gauge symmetry.

Using the Atiyah-Singer index theorem (\ref{8}) and the relations (\ref{3.12}) and (\ref{3.13}), 
we have derived the charge quantization condition (\ref{3.15}). 
This is identical to the charge quantization condition (\ref{1.3}) found by Arafune et al.\cite{AFG}. 
They showed Eq. (\ref{1.3}) by considering continuous mappings from $S^2_{R}$ into 
the unit sphere, $S^2_{\varPhi}$, defined by $\sum_{I=1}^{3} (\varPhi^{I})^{2}=1$. 
According to their analysis, the integer $\frak{n}$ can be geometrically interpreted as    
both the Kronecker index and the Brouwer degree of the mapping $f_{\varPhi}:  
S^2_{R}\rightarrow S^2_{\varPhi}$. The integer $\frak{n}$ is also equal to  
the sum of the Poincar\'{e}-Hopf indices associated with $f_{\varPhi}$.  
Futhermore, $\frak{n}$ can be understood as an integer characterizing the homotopy class 
of $f_{\varPhi}$, or as an element of the homotopy group $\pi_{2} (S^2_{R})=\Bbb Z$. 
Arafune et al. stated that the Kronecker index, the Brouwer degree, 
the sum of the Poincar\'{e}-Hopf indices and the homotopy class are all equivalent ways  
of characterizing $f_{\varPhi}$.

By contrast, we have shown Eq. (\ref{1.3}) without referring to the mapping $f_{\varPhi}$.  
In fact, no mappings like $f_{\varPhi}$ were considered in the proof of Eq. (\ref{8}).  
Also, what we did to derive Eq. (\ref{3.15}) from Eq. (\ref{8}) was only to examine   
the number of chirality zero-modes of the Dirac operator (\ref{3}). 
In this sense, our approach to showing Eq. (\ref{1.3}) is new, and as a result,     
another interpretation of $\frak{n}$ turns out to be possible: {\em The integer $\frak{n}$ 
can be interpreted as half of the index $2(\frak{n}_{++}-\frak{n}_{+-})$ 
of the Dirac operator} (\ref{3}). 
[In fact, we set $\frak{n}=\frak{n}_{++}-\frak{n}_{+-}$ above Eq. (\ref{3.15})]. 
If we take Eq. (\ref{1.3}) as given by other approaches, 
the argument presented in this paper can be understood as an illustration or 
verification of the Atiyah-Singer index theorem in a two-dimensional $SO(3)$ YMH 
system.

In the $U(1)$ gauge theory with a monopole background, the condition corresponding to 
Eq. (\ref{3.14}) is found from the Atiyah-Singer index theorem in two dimensions  
to be\footnote{In Ref.~\citen{DK},  
it was shown that Eq. (\ref{4.1}) reduces to 
the Dirac quantization condition $eg=n/2$ $(n\in \Bbb Z)$ or 
the Schwinger quantization condition $eg=n$ $(n\in \Bbb Z)$, according to  
the choice of the gauge potential.}\cite{DK} 
\begin{align}
\frak{n}_{+}-\frak{n}_{-}=2eg \,. 
\label{4.1}
\end{align}
Here, $\frak{n}_{+}$ ($\frak{n}_{-}$) denotes the number of positive (negative) chirality  
zero-modes of the Dirac operator in the $U(1)$ gauge theory.  It should be noted  
that the right-hand side of this condition is twice that of Eq. (\ref{3.14}).  
This is due to the fact that in the $SO(3)$ YMH theory, 
there exist twice as many chirality zero-modes as in the $U(1)$ gauge theory, 
as may be seen from Eqs. (\ref{3.12}) and (\ref{3.13}).  
The difference between Eqs. (\ref{3.14}) and (\ref{4.1})  
explains, in terms of the Atiyah-Singer index theorem,  
why the charge quantization condition in the $SO(3)$ YMH theory, 
$eg=\frak{n}$ $(\frak{n}\in \Bbb Z)$, 
is different from the Dirac quantization condition, $eg=n/2$ $(n\in \Bbb Z)$,  
by a factor of two.

In the appendix, 
we see that the $SO(3)$ monopole configurations
can be reduced to either Abelian monopole 
configurations of Dirac type or those of Schwinger type, depending on 
the choice of a constant contained in the monopole potential. 
Along the line of the present argument, these two types of configurations can be treated 
in a unified manner without a careful treatment of the string singularity  
in the monopole potential. 
Actually, as can be seen in the appendix, 
both the Dirac and Schwinger charge quntization conditions 
for the $SO(3)$ YMH system can be derived by formally 
counting the number of zero-modes of the Dirac operator.

We have applied the Atiyah-Singer index theorem (\ref{8}) to only 
a particular case in which $S^{2}_{R}$ is chosen as $\mathcal{M}$ and 
$SO(3)$ non-Abelian monopoles are assumed to exist in the system. 
As a future subject of study, 
it would be interesting to consider applications of 
the theorem (\ref{8}) to other physical systems   
in which $\mathcal{M}$ has a non-trivial topology.  
In  these applications,  
new relations other than Eq. (\ref{3.15}) may be found from Eq. (\ref{8}).  
It is also of interest to generalize the Atiyah-Singer index theorem 
to the $SU(N)$ YMH system in two dimensions.   
In the presence of $SU(N)$ non-Abelian monopoles,\cite{GNO}\tocite{Shn} 
such a generalized theorem should provide an analog of the condition (\ref{3.15}).  
We hope to address these issues in the future.

\section*{Acknowledgements}
The author would like to thank Prof. K. Fujikawa for his encouragement and useful 
comments. 
This work was supported in part by the Nihon University Research Grant (No. 06-069).

\appendix
\section{}

This appendix gives an illustration of the charge quantization condition (\ref{3.15})  
using  concrete forms of the $SO(3)$ monopole configurations. 

Let us consider the Yang-Mills connection $A$ and the 
scalar field $\Phi$ defined by 
\begin{align}
A=& \: k( \sin n\phi \,d\theta +n \sin\theta \cos\theta \cos n\phi \,d\phi) \tau_{1}
\nonumber 
\\
& \,+k( \cos n\phi \,d\theta -n \sin\theta \cos\theta \sin n\phi \,d\phi) \tau_{2} 
\nonumber 
\\ 
&  \,-kn \sin^{2} \theta \, d\phi \tau_{3} \,, 
\label{A1} 
\\ 
\varPhi=& \:  \sin\theta \cos n\phi \,\tau_{1} -\sin\theta \sin n\phi \,\tau_{2}
+\cos\theta \,\tau_{3} \,, 
\label{A2}
\end{align}
where $k$ is a real constant and $n$ must be an integer to insure 
the single-valuedness of $A$ and $\varPhi$ 
under a $2\pi$ rotation in $\phi$. 
Up to $k$, the expressions (\ref{A1}) and (\ref{A2}) 
are essentially the same as those given by Bais\cite{Bai,Shn}.

The curvature two-form of $A$ is found from Eq. (\ref{A1}) to be 
\begin{align}
F& =\frac{1}{2} F_{\alpha\beta} dq^{\alpha}  dq^{\beta} 
= dA -i\frac{e}{2} A A 
\nonumber \\ 
& =-2kn \bigg( 1+\frac{ke}{2} \bigg) \varPhi \sin\theta \,d\theta d\phi \,. 
\label{A3}
\end{align}
From Eqs. (\ref{A1}) and (\ref{A2}),  
covariant differentiation of $\varPhi$ is obtained as   
\begin{align}
D\varPhi =d\varPhi -i \frac{e}{2} [A, \varPhi ] = (1+ ke) d\varPhi \,, 
\label{A4}
\end{align}
which leads to    
\begin{align}
D\varPhi  D\varPhi =-2in (1+ke)^{2} \varPhi \sin\theta \,d\theta d\phi \,. 
\label{A5}
\end{align}
From the combination of Eqs. (\ref{A3}) and (\ref{A5}) given by 
\begin{align}
F+ \frac{i}{2e} D\varPhi D\varPhi = \frac{n}{e} \varPhi \sin\theta \,d\theta d\phi \,, 
\label{A6} 
\end{align}
the 't Hooft tensor written in terms of differential forms is found to be   
\begin{align}
\mathcal{F} & =\frac{1}{2} \mathcal{F}_{\alpha\beta} dq^{\alpha}  dq^{\beta} 
= \frac{1}{2}\, {\rm tr} \bigg( \varPhi  
F+\frac{i}{2e} \varPhi D \varPhi D \varPhi  \bigg) 
\nonumber \\ 
&= \frac{n}{e} \sin\theta \,d\theta d\phi \,. 
\label{A7}
\end{align}
Consequently, the conserved magnetic charge (\ref{3.1}) is obtained as    
\begin{align}
g= \frac{1}{4\pi} \int_{S^2_R} \mathcal{F} =\frac{n}{e} \,. 
\label{A8}
\end{align}
Thus, Eqs. (\ref{A1}) and (\ref{A2}) are shown to be 
configurations of $SO(3)$ monopoles with magnetic charge $n/e$ ($n\in {\Bbb Z}$). 
Equation (\ref{A8}) reduces to Eq. (\ref{1.3}) with the identification $\frak{n}=n$. 
Although each of Eqs. (\ref{A3}) and (\ref{A5}) depends on $k$, 
a particular combination (\ref{A6}), and hence Eqs. (\ref{A7}) and (\ref{A8}),   
are independent of $k$. This remarkable fact implies that the 't Hooft tensor and  
the magnetic charge obtained here are actually determined only by the scalar field (\ref{A2}); 
they are independent of the Yang-Mills connection (\ref{A1}). 
In Ref.~\citen{AFG}, Arafune et al. showed that the magnetic charge in the $SO(3)$ YMH theory is 
completely specified in terms of a triplet of Higgs fields. 
The result of our analysis is thus consistent with their statement.

From the scalar field  (\ref{A2}), 
the matrix $v$ satisfying Eq. (\ref{1}) is determined to be 
\begin{align}
v(\theta, \phi) =\left(\begin{array}{cc}
e^{i(n+\bar{n}) \phi} \cos\dfrac{\theta}{2}  & -e^{-i\bar{n}\phi} \sin\dfrac{\theta}{2} \\
e^{i\bar{n}\phi} \sin\dfrac{\theta}{2}  &  e^{-i(n+\bar{n}) \phi} \cos\dfrac{\theta}{2} \\
          \end{array} \right) , 
\label{A9}
\end{align}
where $\bar{n}$ is a real constant. Let us recall Eq. (\ref{3.3}). 
Because $\varphi_{0}^{t,s}$ and $\tilde{\varphi}_{0}^{t,s}$ are spinor fields, 
they change sign under a $2\pi$ rotation in $\phi$.   
Accordingly, as seen from Eq. (\ref{3.3}),  
$v$ must be single-valued under the same rotation, i.e. 
$v(\theta, \phi+2\pi)=v(\theta, \phi)$. 
This condition requires that $\bar{n}$ be an integer, as is $n$. 
Substituting Eqs. (\ref{A1}) and (\ref{A9}) into Eq. (\ref{23.5}), we obtain 
\begin{align}
\tilde{A}=
& \: -\bigg(k+\frac{1}{e} \bigg) 
( \sin l\phi \,d\theta -n \sin\theta \cos l \phi \,d\phi) \tau_{1}
\nonumber \\
& \,+\bigg(k+\frac{1}{e} \bigg) 
( \cos l\phi \,d\theta +n \sin\theta \sin l\phi \,d\phi) \tau_{2} 
\nonumber \\ 
&  \,-\frac{1}{e} (l +n \cos\theta ) d\phi \tau_{3} \,, 
\label{A10}
\end{align}
where $l\equiv n+2\bar{n}$.  The third component, $\tilde{A}^{3}$, is immediately 
read from (\ref{A10}) as    
$\tilde{A}^{3} = -e^{-1} (l +n \cos\theta ) d\phi 
=\tilde{A}^{3}_{\alpha} dq^{\alpha}$, with  
\begin{subequations}
\label{A11}
\begin{align}
&\tilde{A}^{3}_{1}=0 \,, 
\label{A11a}
\\
&\tilde{A}^{3}_{2}=-\frac{1}{e} (l +n \cos\theta )  \,.
\label{A11b}
\end{align}
\end{subequations}
Then, from Eq. (\ref{3.5c}), it follows that  
$\hat{A}^{3}_{2}=-e^{-1} (l +n \cos\theta )$, and hence    
Eq. (\ref{3.10}) takes the form   
\begin{align}
\left[\, \frac{d}{d\theta} + \frac{1}{2} (1-stn) \cot\theta 
-s \bigg( m+\frac{tl}{2} \bigg) \csc \theta\, \right]\! \hat{v}^{t,s}_{m} =0 \,. 
\label{A12}
\end{align}
This equation can readily be solved as 
\begin{align}
\hat{v}^{t,s}_{m}(\theta)=\hat{c}_{m}^{t,s} 
\biggl(\sin\frac{\theta}{2}\biggr)^{p^{t.s}_{m}} 
\biggl(\cos\frac{\theta}{2}\biggr)^{q^{t,s}_{m}} , 
\label{A13}
\end{align}
where $\hat{c}_{m}^{t,s}$ is an appropriate constant, and 
\begin{align}
p^{t,s}_{m} &\equiv sm+\frac{1}{2} \{ st(n+l)-1 \} \,,  
\label{A14} 
\\ 
q^{t,s}_{m} &\equiv -sm+\frac{1}{2}  \{ st(n-l)-1 \} \,. 
\label{A15}
\end{align}
The solution $\hat{v}^{t,s}_{m}$ diverges at neither $\theta=0$ nor $\pi$ 
if and only if $p^{t,s}_{m}, q^{t,s}_{m}\geq 0$.  In this case,  
$\hat{v}^{t,s}_{m}$ is normalizable with respect to the usual $L^2$ norm\cite{DK}. 
This fact enables us to choose  $\hat{c}_{m}^{t,s}$ in the case 
$p^{t,s}_{m}, q^{t,s}_{m}\geq 0$ to be a normalization constant of $\hat{v}_{m}^{t,s}$. 
The conditions $p^{t,s}_{m}, q^{t,s}_{m}\geq 0$ necessary for $\hat{v}^{t,s}_{m}$ to  
be regular can together be written as  
\begin{align}
-\frac{1}{2} \{ st(n+l)-1 \} \leq sm \leq  \frac{1}{2}  \{ st(n-l)-1 \} \,. 
\label{A16}
\end{align}
From Eq. (\ref{A16}), it follows that 
$-\{ st(n+l)-1 \} \leq   \{ st(n-l)-1 \}$, which can be simplified as $stn \geq 1$.  
This condition implies that if $n$ is a positive 
integer, $n \in {\Bbb Z}^{+}$, then $st=+$. Hence, when $n \in {\Bbb Z}^{+}$, 
there exist no regular solutions $\hat{v}^{+-}_{m}$ or $\hat{v}^{-+}_{m}$, 
and thus $\frak{n}_{+-}=\frak{n}_{-+}=0$.  
Similarly, if $n$ is a negative integer, $n \in {\Bbb Z}^{-}$, then 
the condition $stn \geq 1$ implies $st=-$.  
Hence, when $n \in {\Bbb Z}^{-}$, there exist no regular solutions $\hat{v}^{++}_{m}$  
or $\hat{v}^{--}_{m}$, and thus that $\frak{n}_{++}=\frak{n}_{--}=0$.  
It is now obvious that when $n=0$, there exist no 
regular solutions, and hence $\frak{n}_{t,s}=0$  $(t,s=+,-)$. 
This illustrates the Lichnerowicz vanishing theorem\cite{Lic}.

In the following, we consider the two cases $l=n$ and $l=0$ in particular, 
because in these cases, it is easy to count the number of  
regular solutions $\hat{v}^{++}_{m}$ and $\hat{v}^{--}_{m}$ for $n \in {\Bbb Z}^{+}$ 
and the number of regular solutions 
$\hat{v}^{+-}_{m}$ and $\hat{v}^{-+}_{m}$ for $n \in {\Bbb Z}^{-}$.

\subsection{The case $l=n$} 

In the case $l=n$, $\hat{A}^{3}_{2}$ reduces to a monopole potential of Dirac type,  
$\hat{A}^{3}_{2}=-e^{-1} n(1 + \cos\theta )$, and Eq. (\ref{A16}) reads 
\begin{align}
-stn+\frac{1}{2} \leq sm \leq  -\frac{1}{2}  \,. 
\label{A17}
\end{align}
First, suppose that $n \in {\Bbb Z}^{+}$, or equivalently $st=+$.   
Then Eq. (\ref{A17}) becomes $-(2n-1)/2 \leq sm \leq -1/2$. 
Because $m$ takes half-integer values, the allowed values of $sm$ are seen  
to be $sm=-1/2, -3/2, \ldots, -(2n-1)/2$.  This implies that the number of regular 
solutions $\hat{v}^{++}_{m}$ and the number of regular solutions $\hat{v}^{--}_{m}$ are 
both $n$, and it follows that $\frak{n}_{++}=\frak{n}_{--}=n$. 
As a result, taking into account the fact that 
$\frak{n}_{+-}=\frak{n}_{-+}=0$ for $n \in {\Bbb Z}^{+}$, we have   
\begin{align}
\frak{n}=\frak{n}_{++}-\frak{n}_{+-}=n\,, \quad n \in {\Bbb Z}^{+} . 
\label{A18}
\end{align}
Next, suppose that $n \in {\Bbb Z}^{-}$, or equivalently $st=-$.   
Then Eq. (\ref{A17}) becomes $(2n+1)/2 \leq sm \leq -1/2$,  
and the allowed values of $sm$ are found to be   
$sm=-1/2, -3/2, \ldots, (2n+1)/2$.  This implies that the number of regular 
solutions $\hat{v}^{+-}_{m}$ and the number of regular solutions $\hat{v}^{-+}_{m}$ are 
both $-n$, and it follows that $\frak{n}_{+-}=\frak{n}_{-+}=-n$. 
Recalling that $\frak{n}_{++}=\frak{n}_{--}=0$ for $n \in {\Bbb Z}^{-}$,  
we have   
\begin{align}
\frak{n}=\frak{n}_{++}-\frak{n}_{+-}=n\,, \quad n \in {\Bbb Z}^{-} . 
\label{A19}
\end{align}
Equations (\ref{A18}) and (\ref{A19}), 
together with the fact that $\frak{n}_{t,s}=0$ for $n=0$, 
are brought together in the form $\frak{n}=n$ with $n\in {\Bbb Z}$.  
Combining this with Eq. (\ref{A8}) leads to $eg=\frak{n}$, with $\frak{n}\in {\Bbb Z}$. 
Thus, the charge quantization condition (\ref{3.15}) is illustrated with the 
monopole configurations (\ref{A1}) and (\ref{A2}). 

If $l=-n$, 
$\hat{A}^{3}_{2}$ reduces to another monopole potential of Dirac type,  
$\hat{A}^{3}_{2}=e^{-1} n(1 -\cos\theta )$. This is merely a mirror image of 
the potential in the case $l=n$. 
Following the same procedure as in the case $l=n$,  
we again obtain the condition $eg=\frak{n}$ with $\frak{n}\in {\Bbb Z}$.

\subsection{The case $l=0$} 

In the case $l=0$, $\hat{A}^{3}_{2}$ reduces to a monopole potential of Schwinger type,  
$\hat{A}^{3}_{2}=-e^{-1} n\cos\theta $, and Eq. (\ref{A16}) reads 
\begin{align}
-\frac{1}{2}stn+\frac{1}{2} \leq sm \leq  \frac{1}{2}stn-\frac{1}{2}  \,. 
\label{A20}
\end{align}
We should note that $n$ here takes only even values, because $l=0$ implies   
$n=-2\bar{n}$ and $\bar{n}$ takes integer values. 

First, suppose that $n$ is a positive even integer,  $n \in 2{\Bbb Z}^{+}$, 
or equivalently $st=+$.  
Then Eq. (\ref{A20}) becomes $-(n-1)/2 \leq sm \leq (n-1)/2$. 
Because $m$ takes half-integer values, the allowed values of $sm$ are seen  
to be $sm=\pm1/2, \pm3/2, \ldots, \pm(n-1)/2$.  This implies that the number of regular 
solutions $\hat{v}^{++}_{m}$ and the number of regular solutions $\hat{v}^{--}_{m}$ are 
both $n$, and it follows that $\frak{n}_{++}=\frak{n}_{--}=n$. 
As a result, taking into account the fact that 
$\frak{n}_{+-}=\frak{n}_{-+}=0$ for $n \in 2{\Bbb Z}^{+}$, we have   
\begin{align}
\frak{n}=\frak{n}_{++}-\frak{n}_{+-}=n\,, \quad n \in 2{\Bbb Z}^{+} . 
\label{A21}
\end{align}
Next, suppose that $n$ is a negative even integer, $n \in 2{\Bbb Z}^{-}$, 
or equivalently $st=-$.  
Then Eq. (\ref{A20}) becomes $(n+1)/2 \leq sm \leq -(n+1)/2$, 
and the allowed values of $sm$ are found to be   
$sm=\pm1/2, \pm3/2, \ldots, \pm(n+1)/2$. 
This implies that the number of regular 
solutions $\hat{v}^{+-}_{m}$ and the number of regular solutions $\hat{v}^{-+}_{m}$ are 
both $-n$, and it follows that $\frak{n}_{+-}=\frak{n}_{-+}=-n$. 
Noting that $\frak{n}_{++}=\frak{n}_{--}=0$ for $n \in 2{\Bbb Z}^{-}$, we have   
\begin{align}
\frak{n}=\frak{n}_{++}-\frak{n}_{+-}=n\,, \quad n \in 2{\Bbb Z}^{-} . 
\label{A22} 
\end{align}
Equations (\ref{A21}) and (\ref{A22}), 
together with the fact that $\frak{n}_{t,s}=0$ for $n=0$, 
are brought together in the form $\frak{n}=n$, 
where $n$ is an even integer,  $n\in 2{\Bbb Z}$.  
Combining this with Eq. (\ref{A8}) leads to $eg=\frak{n}$ with $\frak{n}\in 2{\Bbb Z}$. 
Thus, the charge quantization condition (\ref{3.15}) is illustrated also in the case $l=0$ with the 
configurations (\ref{A1}) and (\ref{A2}), though $n$ here is restricted to even integers.

\subsection{Comments} 

It has been seen that when $l=n$ (or $l=-n$), $n$ can take all integer values, 
while when $l=0$, $n$ can take only even values. 
This fact implies the following:  
If  it is assumed that $n$ in Eqs.  (\ref{A1}) and (\ref{A2}) can take all integer values, 
then it is possible to make gauge transformations that yield the Dirac potentials 
$\hat{A}^{3}_{2}=\mp e^{-1} n(1 \pm\cos\theta )$; in this case, 
the gauge transformation 
that gives the Schwinger potential $\hat{A}^{3}_{2}=-e^{-1} n\cos\theta $ is not allowed. 
By contrast, if $n$ in Eqs.  (\ref{A1}) and (\ref{A2}) is assumed to take only even values,  
the gauge transformation that gives the Schwinger potential is allowed. 
In this case, it is, of course, 
possible to make the gauge transformations that give the Dirac potentials. 
In this way, the allowed gauge transformations are determined by the integer values 
that $n$ takes.

\end{document}